\documentclass[
    ,final            
  ]
  {aipproc}

\layoutstyle{8x11single}

\begin{document}

\title{On the properties of the RHESSI intermediate-duration gamma-ray bursts}

\classification{01.30.Cc, 95.85.Pw, 98.70.Rz}

\keywords{gamma-ray astrophysics, gamma-ray bursts}

\author{Jakub \v{R}\'{i}pa}{address={Charles University, Faculty of Mathematics and Physics, Astronomical Institute,\\ V Hole\v{s}ovi\v{c}k\'ach 2, 180 00 Prague 8, Czech Republic}}

\author{P\'{e}ter Veres}{address={E\"{o}tv\"{o}s Lor\'{a}nd University,
P\'{a}zm\'{a}ny P\'{e}ter s\'{e}t\'{a}ny 1/A, Budapest, Hungary}}

\author{Attila M\'{e}sz\'{a}ros}{address={Charles University, Faculty of Mathematics and Physics, Astronomical Institute,\\ V Hole\v{s}ovi\v{c}k\'ach 2, 180 00 Prague 8, Czech Republic}}

\begin{abstract}
The intermediate-duration gamma-ray bursts (GRBs) identified in the data of the RHESSI satellite are investigated with respect to their spectral lags, peak count rates, redshifts, supernova observations, and star formation rates of their host galaxies. Standard statistical tests like Kolmogorov-Smirnov and Student t-test are used. It is discussed whether these bursts belong to the group of so-called short or long GRBs, or if they significantly differ from both groups.
\end{abstract}

\maketitle

\section{Methods}
\begin{figure}[h]$
\begin{array}{cc}
\includegraphics[trim=15 6 14 17,clip=true,width=7.8cm]{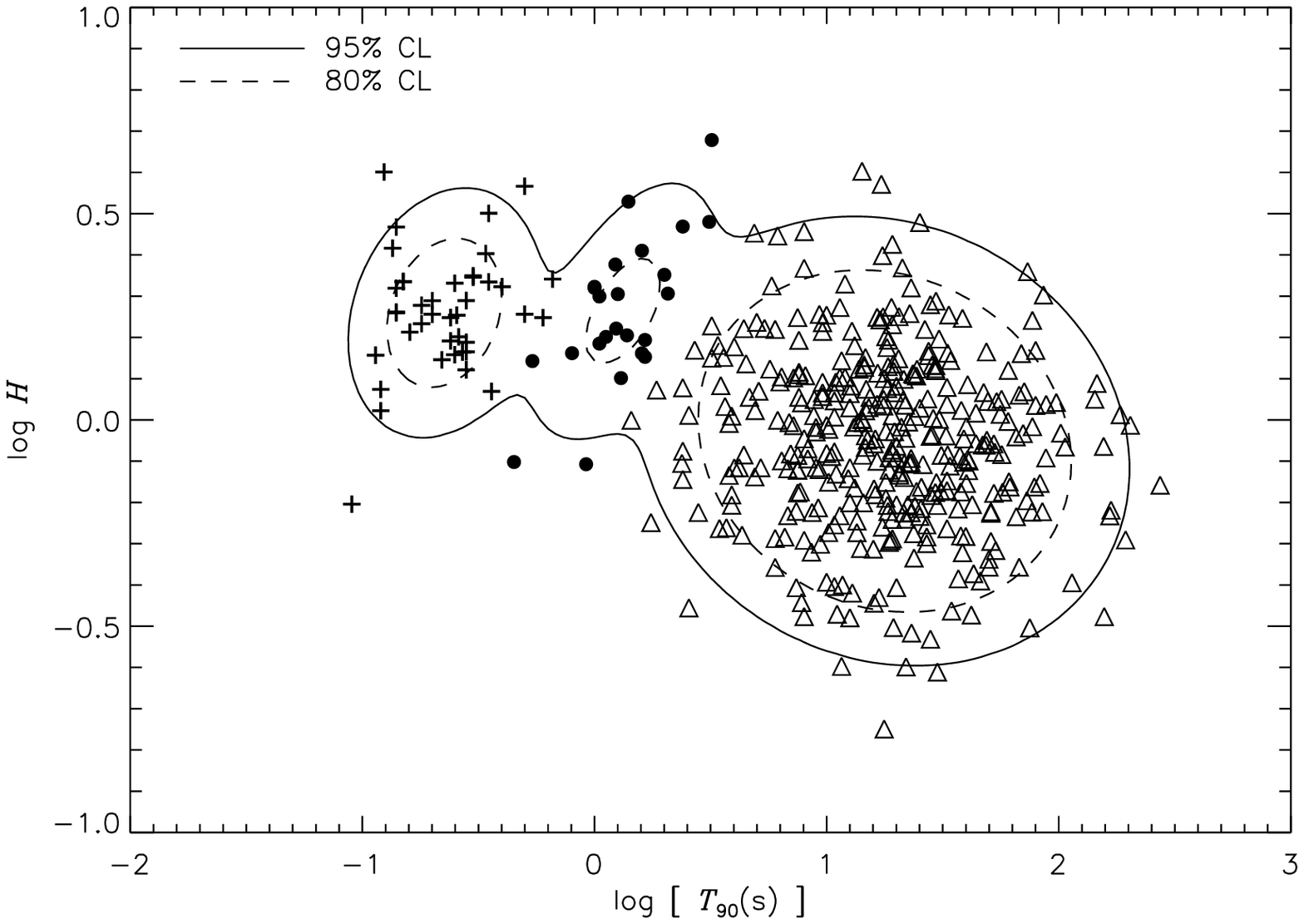}
&
\includegraphics[trim=15 3 14 18,clip=true,width=7.7cm]{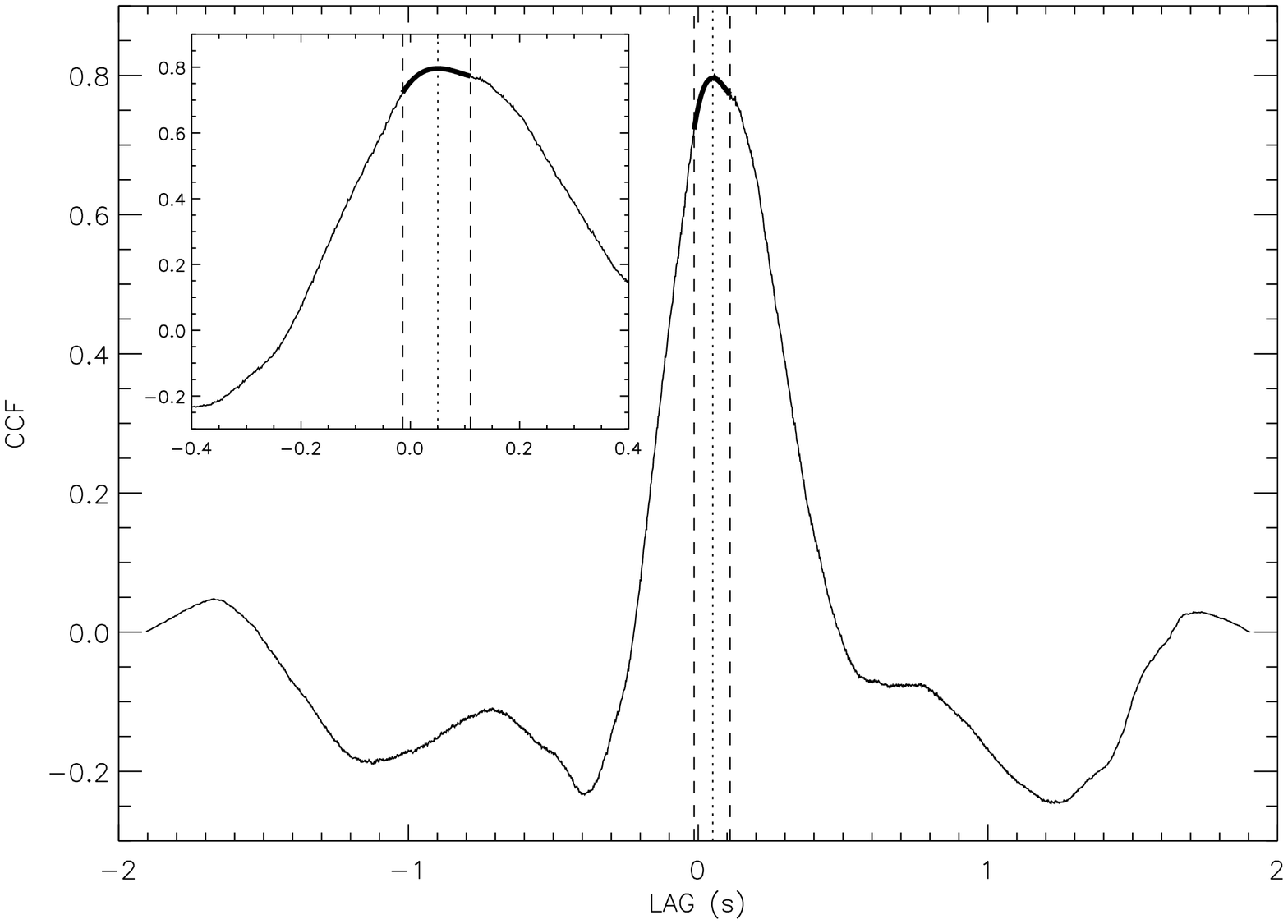}
\end{array}$
\caption{
\emph{Left panel:}
The hardness ratio $H$ vs. $T_{90}$ duration plot with the best ML fit of three bivariate log-normal functions is shown. Different symbols correspond to different GRB groups identified in the article \citet{rip09}. For each GRB the probability of belonging to either short-, intermediate-, or long-duration group is known. The crosses, full circles, and triangles mark GRBs belonging, by the highest probability, to the group of short-, intermediate-, or long-duration bursts, respectively.
\emph{Right panel:}
An example of cross-correlation function of background-subtracted count light-curve (here very bright GRB~060306) at two different energy bands. The inset presents the fit of the CCF peak by the third order polynomial (thick solid curve). The maximum of the polynomial fit (dotted line) was taken as the true spectral lag. The boundaries of the fit are marked with dashed lines.}
\label{fig:1-2}
\end{figure}

In paper \cite{rip09} we identified a group of intermediate-duration GRBs (here Figure~\ref{fig:1-2}, left panel), significant at $\sim3\sigma$ level, by the Maximum Likelihood (ML) ratio test and bivariate fitting of log-normal functions on the hardness ratio-duration plane. Here we focus on the other properties of these bursts and employ the same sample, i.e., the data obtained from the RHESSI satellite between February 14, 2002 and April 25, 2008. The number of the GRBs belonging to the group of, by the highest probability, short, intermediate, and long bursts is 40, 24, 363, respectively.

\begin{figure}[!hp]$
\begin{array}{cc}
\includegraphics[trim=21 6 5 17,clip=true,width=7.8cm]{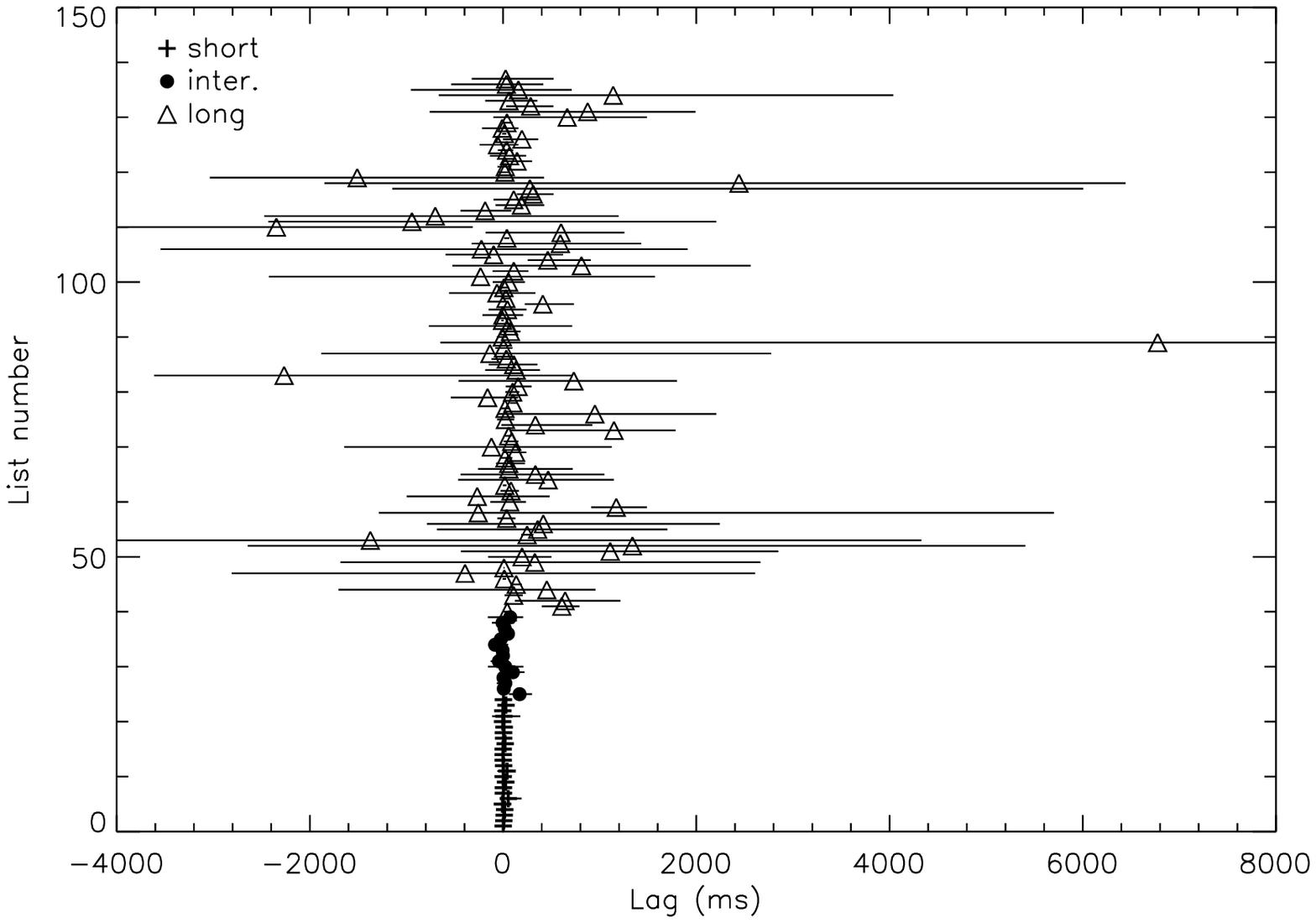}
&
\includegraphics[trim=24 6 5 17,clip=true,width=7.8cm]{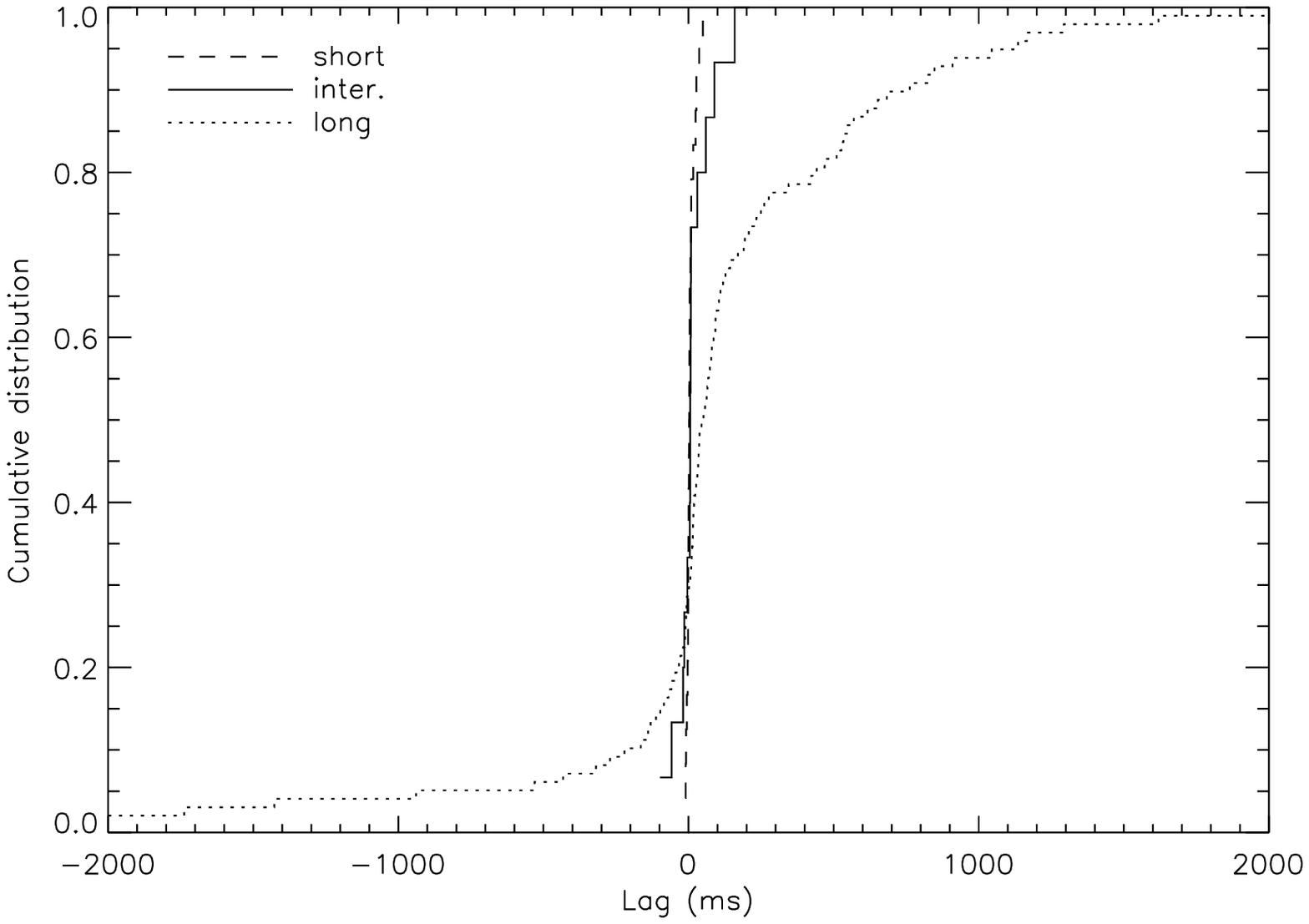}
\end{array}$
\caption{
\emph{Left panel:}
The spectral lags of RHESSI GRBs. Crosses, full circles, and triangles denotes median lags for short-, intermediate-, and long-duration GRBs, respectively. The median lags, for each GRB, were taken from the lags of 1001 synthetic background-subtracted count light-curves obtained by the Monte Carlo (MC) simulations of the measured light-curves that were randomly influenced by the Poissonian noise. The error bars are of more than 95\% CL and composed of a statistical error (0.025 quantile of the lags from the 1001 MC simulations) and the light-curve time resolution. A positive lag means that low-energy counts are delayed.
\emph{Right panel:}
The cumulative distributions of the obtained median lags for the three identified GRB groups. A difference between lags of long- and short- (or intermediate-) duration bursts is clearly seen. The lag distributions of short and intermediate GRBs are similar.}
\label{fig:3-4}
\end{figure}

\begin{figure}[!hp]$
\begin{array}{cc}
\includegraphics[trim=32 6 15 17,clip=true,width=7.8cm]{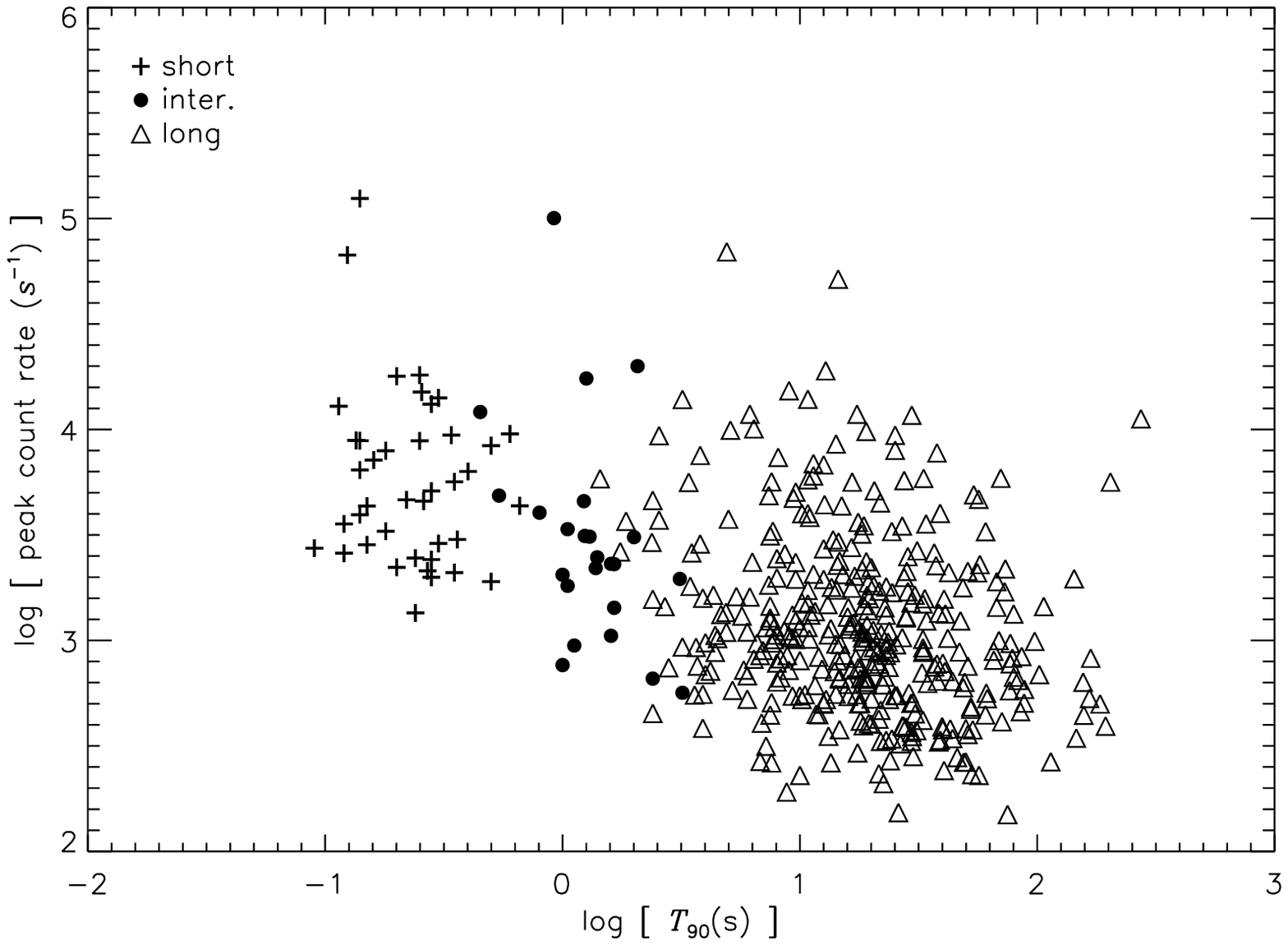}
&
\includegraphics[trim=24 6 15 17,clip=true,width=7.8cm]{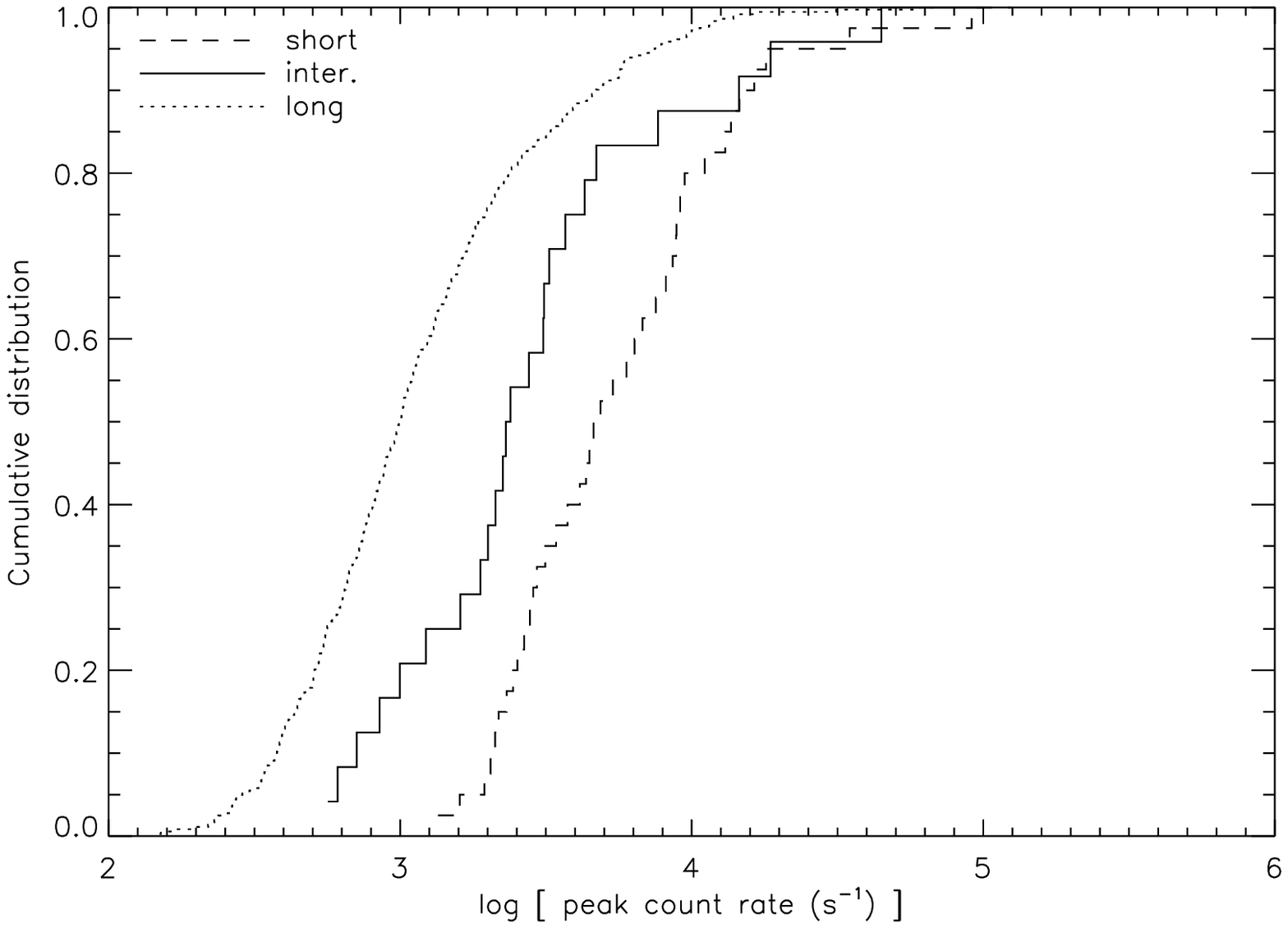}
\end{array}$
\caption{
\emph{Left panel:}
Logarithmic peak count rates as a function of logarithmic $T_{90}$ durations for the three GRB groups, identified by the analysis of the hardnesses and durations, are displayed.
\emph{Right panel:}
Cumulative distributions of peak count rates for the short-, intermediate-, and long-duration RHESSI GRBs.}
\label{fig:5-6}
\end{figure}

We evaluated spectral lags, similarly to \cite{nor02,fol09}, by fitting of the peak of the cross-correlation function (CCF) of the background-subtracted count light-curves at two channels $400-1500$\,keV and $25-120$\,keV (see an example in Figure~\ref{fig:1-2}, right panel). Then we applied Kolmogorov-Smirnov (K-S) and Student t-test to find whether the group of intermediate-duration GRBs is more similar to the group of short or long bursts. In some cases the evaluation of the lag was impossible, because of the high noise level of the CCF. Therefore the sample of the lags is smaller than the overall number of the RHESSI bursts. Results are shown in Figure~\ref{fig:3-4} and summarized in Table~\ref{tab:1-2}. Next we focused on the count peak rates. Figure~\ref{fig:5-6} presents the peak count rates vs. $T_{90}$ durations and the cumulative distributions of the all three identified groups. The K-S and t-test probabilities were calculated and results are presented in Table~\ref{tab:3-4}.

\section{Results and Conclusions}
The K-S test applied on the spectral lag distributions of the short- and intermediate-duration bursts gives the K-S probability 20.7\,\% and K-S distance 0.33. The tabular critical value of the K-S distance at the significance level $\alpha = 5$\,\% and the given number of the elements (24 and 15) is 0.45. Therefore the null hypothesis that the lags of the short and intermediate bursts are drawn from the same distribution cannot be rejected at the 5\,\% significance level.

On the other hand, the same test applied on the lags of the intermediate vs. long (and short vs. long) bursts gives the K-S probability 1.47\,\% (0.01\,\%) and the K-S distance 0.42 (0.49). The tabular critical value of the K-S distance at the 5\,\% significance level for the number of the intermediate and long bursts (15 and 98) is 0.38. Therefore the hypothesis that the lags of the intermediate and long bursts are drawn from the same distribution can be rejected at this 5\,\% significance level.

Similarly, it can be rejected, at the 5\,\% significance level, that the lags of short and long bursts are drawn from the same distribution. The critical K-S distance is in this case 0.31.

\begin{table}[h]$
\begin{array}{cc}
\begin{tabular}{ccccc}
\hline
   \tablehead{1}{c}{b}{Groups\\}
 & \tablehead{1}{c}{b}{K-S dist.\\}
 & \tablehead{1}{c}{b}{K-S prob.\\(\%)}
 & \tablehead{1}{c}{b}{t\\}
 & \tablehead{1}{c}{b}{t-test prob.\\(\%)}\\
\hline
inter.-short & 0.33 & 20.7          & -1.02 & 31.6  \\
inter.-long  & 0.42 & \textbf{1.47} &  0.68 &  49.9 \\
short-long   & 0.49 & \textbf{0.01} &  0.94 &  35.0 \\
\hline
\end{tabular}
&
\begin{tabular}{cccc}
\hline
   \tablehead{1}{c}{b}{Group\\}
 & \tablehead{1}{c}{b}{Mean lag\\(ms)}
 & \tablehead{1}{c}{b}{Median lag\\(ms)}
 & \tablehead{1}{c}{b}{Std. dev.\\(ms)}\\
\hline
short & -7.7   & -1.9  & 16.2  \\
inter.  & -22.5  & -5.9  & 68.8  \\
long  & -181.0 & -50.8 & 902.3 \\
\hline
\end{tabular}
\end{array}$
\caption{
\emph{Left part:}
Results from the Kolmogorov-Smirnov test of equality of distributions of the spectral lags and Student t-test of equality of the mean lags for different GRB groups are presented. The K-S distance, and the K-S probability is mentioned. The null hypothesis that two samples are drawn from the same distribution can be rejected only at a higher or equal significance level than the value of the K-S probability. Also the t-test value and the probability that two samples have equal mean lags are written. Probabilities lower than 5\,\% are highlighted by the boldface.
\emph{Right part:}
The mean, median and standard deviations of the lags are shown.}
\label{tab:1-2}
\end{table}

From the results of the K-S test applied on the peak count rates it follows that the distributions of the peak count rates are different over all three groups. The t-test is not persuasive for both cases; spectral lags and peak count rates.

\begin{figure}
\includegraphics[trim=5 47 6 62,clip=true,width=8.0cm]{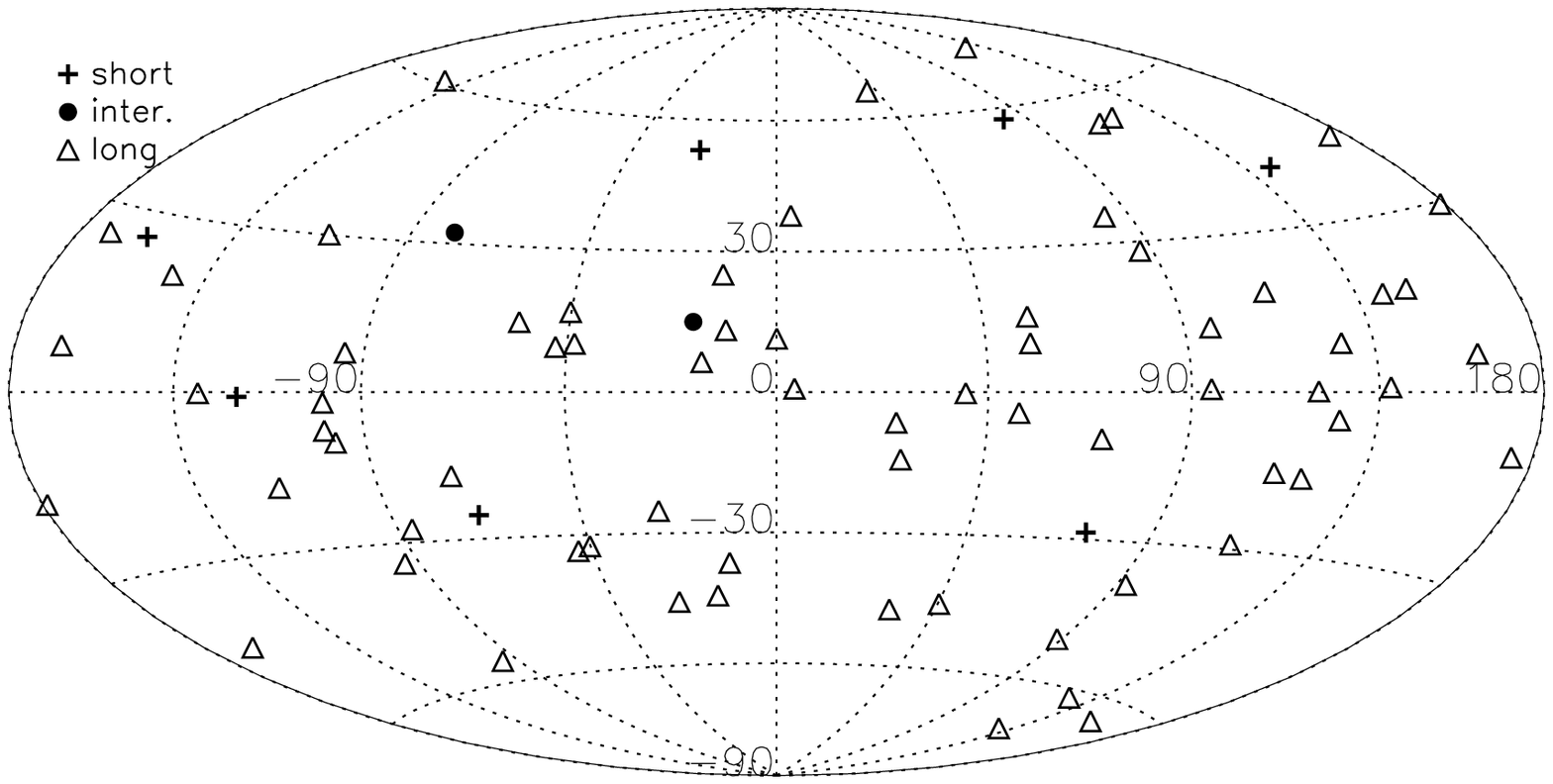}
\caption{Sky distribution of the three RHESSI GRB groups in the Hammer-Aitoff projection with Galactic coordinates (longitude is horizontal).}
\label{fig:7}
\end{figure}

Figure~\ref{fig:7} displays the sky distribution of all localized bursts in our sample. Unfortunately there are only two well localized intermediate events, therefore any statistical analysis is meaningless.

Concerning supernova observations (SN), we found that all GRBs with underlying or possible underlying SN have, by RHESSI, long duration and there is no SN observation for any RHESSI intermediate-duration burst. This finding is similar to the fact that, in general, there is no SN-GRB connection for short-duration bursts associated with the model of the merger of two compact astrophysical objects.

We also tried to find the star formation rates of the host galaxies in the published articles (e.g., \cite{ber09,lev10}) of the intermediate bursts and compare them with the hosts of the short and long bursts. However, we were unsuccessful in finding the measurements of the star formation rates for the RHESSI intermediate-duration bursts.

For the redshift measurements of the bursts observed by RHESSI, the average value for the short busts is $0.49\pm0.69$ and for the long ones $1.16\pm0.33$. Unfortunately no redshift measurements are available for the intermediate bursts.

\begin{table}[!hp]$
\begin{array}{cc}
\begin{tabular}{ccccc}
\hline
   \tablehead{1}{c}{b}{Groups\\}
 & \tablehead{1}{c}{b}{K-S dist.\\}
 & \tablehead{1}{c}{b}{K-S prob.\\(\%)}
 & \tablehead{1}{c}{b}{t\\}
 & \tablehead{1}{c}{b}{t-test prob.\\(\%)}\\
\hline
inter.-short &  0.36 & \textbf{3.1}   & -2.43  &  \textbf{1.8}  \\
inter.-long  &  0.48 & \textbf{3E-3}  &  4.56  &  \textbf{7E-4} \\
short-long &  0.72 & \textbf{2E-15} &  9.86  &  \textbf{1E-18}  \\
\hline
\end{tabular}
&
\begin{tabular}{cccc}
\hline
   \tablehead{1}{c}{b}{Group\\}
 & \tablehead{1}{c}{b}{Mean\\log $F$\,(s$^{-1}$)}
 & \tablehead{1}{c}{b}{Median\\log $\,F$(s$^{-1}$)}
 & \tablehead{1}{c}{b}{Std. dev.\\}\\
\hline
short &   3.76   &   3.71   &    0.41  \\
inter.  &   3.48   &   3.40   &    0.51  \\
long  &   3.06   &   3.00   &    0.43  \\
\hline
\end{tabular}
\end{array}$
\caption{
\emph{Left part:}
Results of the K-S and t-tests applied on the logarithmic peak count rates (log $F$). The shortcuts have the same meaning as in the Table~\ref{tab:1-2}.
\emph{Right part:}
The mean, median and standard deviations of the logarithmic peak count rates. Probabilities lower than 5\,\% are again emphasize by the boldface.}
\label{tab:3-4}
\end{table}

\begin{theacknowledgments}
This study was supported by the Grant Agency of the Czech Republic, grants No. P209/10/0734, No. 205/08/H005, by the Research Program MSM0021620860 of the Ministry of Education of the Czech Republic, by the project SVV 261301 of the Charles University in Prague, and by the OTKA grant No. K77795.
\end{theacknowledgments}

\end{document}